\documentclass[12pt]{article}

\usepackage{axodraw}
\usepackage{epsf}
\usepackage{rotating}

\newcommand{\beq}{\begin{equation}}
\newcommand{\eeq}{\end{equation}}
\newcommand{\bea}{\begin{eqnarray}}
\newcommand{\eea}{\end{eqnarray}}
\newcommand{\bed}{\begin{displaymath}}
\newcommand{\eed}{\end{displaymath}}

\newcommand{\tga}{{\rm tg}\alpha}
\newcommand{\tgb}{{\rm tg}\beta}
\newcommand{\stgb}{{\rm tg}^2\beta}
\newcommand{\sgl}{\tilde{g}}

\newcommand{\st}{\tilde t}
\newcommand{\ssb}{\tilde b}
\newcommand{\sq}{\tilde q}

\newcommand{\gl}{\tilde g}

\newcommand{\gev}{~\mbox{GeV}}

\newcommand{\lsim}{\raisebox{-0.13cm}{~\shortstack{$<$ \\[-0.07cm] $\sim$}}~}

\catcode`\@=11
\def\@citex[#1]#2{\if@filesw\immediate\write\@auxout{\string\citation{#2}}\fi
  \def\@citea{}\@cite{\@for\@citeb:=#2\do
    {\@citea\def\@citea{,\penalty\@m}\@ifundefined
       {b@\@citeb}{{\bf ?}\@warning
       {Citation `\@citeb' on page \thepage \space undefined}}%
\hbox{\csname b@\@citeb\endcsname}}}{#1}}

\def\citer{\@ifnextchar [{\@tempswatrue\@citexr}{\@tempswafalse\@citexr[]}}
\def\@citexr[#1]#2{\if@filesw\immediate\write\@auxout{\string\citation{#2}}\fi
  \def\@citea{}\@cite{\@for\@citeb:=#2\do
    {\@citea\def\@citea{--\penalty\@m}\@ifundefined
       {b@\@citeb}{{\bf ?}\@warning
       {Citation `\@citeb' on page \thepage \space undefined}}%
\hbox{\csname b@\@citeb\endcsname}}}{#1}}
\catcode`\@=12

\begin{document}

\renewcommand{\thefootnote}{\fnsymbol{footnote}}
\setcounter{page}{0}

\begin{titlepage}

\vskip-1.0cm

\begin{flushright}
PSI--PR--05--01 \\
hep-ph/0501164
\end{flushright}

\begin{center}
{\large \sc Associated Higgs Boson Production with heavy Quarks} \\[0.5cm]
{\large \sc in $e^+e^-$ Collisions: SUSY--QCD
Corrections}\footnote{Supported in part by the Swiss Bundesamt f\"ur
Bildung und Wissenschaft.}\\
\end{center}

\vskip 1.cm
\begin{center}
{\sc Petra H\"afliger$^{1,2}$ and Michael Spira$^2$}

\vskip 0.8cm

\begin{small} 
{\it \small
$^1$ Institute for Particle Physics, ETH Z\"urich, CH-8093 Z\"urich,
Switzerland \\
$^2$ Paul Scherrer Institut, CH-5232 Villigen PSI, Switzerland}
\end{small}
\end{center}

\vskip 2cm

\begin{abstract}
\noindent
The processes $e^+e^-\to t\bar{t}/b\bar b + \mbox{Higgs}$ allow to
measure the Yukawa couplings between Higgs bosons and heavy quarks in
supersymmetric theories. The complete set of next-to-leading order
SUSY--QCD corrections to the cross sections of these processes have been
determined in the minimal supersymmetric extension of the Standard
Model. They turn out to be ${\cal O}(10-20\%)$ and thus important for
future linear $e^+e^-$ colliders.
\end{abstract}

\end{titlepage}

\renewcommand{\thefootnote}{\arabic{footnote}}

\setcounter{footnote}{0}

\section{Introduction}
%        ============
The Higgs mechanism is a cornerstone of the Standard Model (SM) and its
supersymmetric extensions \cite{hi64}. In the minimal supersymmetric
extension of the SM (MSSM) two isospin Higgs doublets have to be
introduced in order to generate masses of up- and down-type fermions
\cite{twoiso}. After electroweak symmetry breaking three of the eight
degrees of freedom are absorbed by the $Z$ and $W$ gauge bosons,
implying the existence of five elementary Higgs particles. These consist
of two CP-even neutral (scalar) particles $h,H$, one CP-odd neutral
(pseudoscalar) particle $A$ and two charged bosons $H^\pm$.  At leading
order the MSSM Higgs sector is fixed by two independent input parameters
which are usually chosen to be the pseudoscalar Higgs mass $M_A$ and
$\tgb=v_2/v_1$, the ratio of the two vacuum expectation values.
Including the one-loop and dominant two-loop corrections the upper bound
of the light scalar Higgs mass is $M_h\lsim 135$ GeV \cite{mssmrad}. The
couplings of the various neutral Higgs bosons to fermions and gauge
bosons, normalized to the SM Higgs couplings, are listed in
Table~\ref{tb:hcoup}, where the angle $\alpha$ denotes the mixing angle
of the scalar Higgs bosons $h,H$. An important property of the bottom
Yukawa couplings is their enhancement for large values of $\tgb$, while
the top Yukawa couplings are suppressed for large $\tgb$.  This implies
that Higgs radiation off bottom quarks plays a significant role at
linear $e^+e^-$ colliders in the large $\tgb$ regime, while Higgs
radiation off top quarks is relevant for small and moderate values of
$\tgb$.
\begin{table}[hbt]
\renewcommand{\arraystretch}{1.5}
\begin{center}
\begin{tabular}{|lc||ccc|} \hline
\multicolumn{2}{|c||}{$\phi$} & $g^\phi_u$ & $g^\phi_d$ &  $g^\phi_V$ \\
\hline \hline
SM~ & $H$ & 1 & 1 & 1 \\ \hline
MSSM~ & $h$ & $\cos\alpha/\sin\beta$ & $-\sin\alpha/\cos\beta$ &
$\sin(\beta-\alpha)$ \\ & $H$ & $\sin\alpha/\sin\beta$ &
$\cos\alpha/\cos\beta$ & $\cos(\beta-\alpha)$ \\
& $A$ & $ 1/\tgb$ & $\tgb$ & 0 \\ \hline
\end{tabular}
\renewcommand{\arraystretch}{1.2}
\caption[]{\label{tb:hcoup} \it Higgs couplings in the MSSM to fermions
and gauge bosons [$V=W,Z$] relative to the SM couplings.}
\end{center}
\end{table}

For the computation of the SUSY-QCD corrections we need the Higgs
couplings to stop and sbottom squarks in addition, as well as the squark
couplings to the photon and $Z$ boson. The scalar superpartners $\tilde
f_{L,R}$ of the left- and right-handed fermion components mix with each
other. The mass eigenstates $\tilde f_{1,2}$ of the sfermions
are related to the current eigenstates $\tilde f_{L,R}$ by mixing angles
$\theta_f$,
\begin{eqnarray}
\tilde f_1 & = & \tilde f_L \cos\theta_f + \tilde f_R \sin \theta_f
\nonumber \\
\tilde f_2 & = & -\tilde f_L\sin\theta_f + \tilde f_R \cos \theta_f \, .
\label{eq:sfmix}
\end{eqnarray}
The mass matrix of the sfermions in the left-right-basis is given by
\cite{mssmbase}\footnote{For simplicity, the $D$-terms have been
absorbed in the sfermion-mass parameters $M_{\tilde f_{L/R}}^2$.}
\begin{equation}
{\cal M}_{\tilde f} = \left[ \begin{array}{cc}
M_{\tilde f_L}^2 + m_f^2 & m_f (A_f-\mu r_f) \\
m_f (A_f-\mu r_f) & M_{\tilde f_R}^2 + m_f^2
\end{array} \right] \, ,
\label{eq:sqmassmat}
\end{equation}
with the parameters $r_d = 1/r_u = \tgb$. The parameters $A_f$
denote the trilinear scalar couplings of the soft supersymmetry breaking
part of the Lagrangian. The mixing angles acquire the form
\begin{equation}
\sin 2\theta_f = \frac{2m_f (A_f-\mu r_f)}{M_{\tilde f_1}^2 - M_{\tilde
f_2}^2}
~~~,~~~
\cos 2\theta_f = \frac{M_{\tilde f_L}^2 - M_{\tilde f_R}^2}{M_{\tilde
f_1}^2
- M_{\tilde f_2}^2} \, ,
\end{equation}
and the masses of the squark mass eigenstates are given by
\begin{equation}
M_{\tilde f_{1,2}}^2 = m_f^2 + \frac{1}{2}\left[ M_{\tilde f_L}^2 +
M_{\tilde f_R}^2 \mp \sqrt{(M_{\tilde f_L}^2 - M_{\tilde f_R}^2)^2 +
4m_f^2 (A_f - \mu r_f)^2} \right] \, .
\end{equation}
Since the mixing angles are proportional to the masses of the ordinary
fermions, mixing effects are only important for the third-generation
sfermions. The neutral Higgs couplings to sfermions read \cite{DSUSY}
\begin{eqnarray}
g_{\tilde f_L \tilde f_L}^\phi & = & m_f^2 g_1^\phi + M_Z^2 (I_{3f}
- e_f\sin^2\theta_W) g_2^\phi \nonumber \\
g_{\tilde f_R \tilde f_R}^\phi & = & m_f^2 g_1^\phi + M_Z^2
e_f\sin^2\theta_W
g_2^\phi \nonumber \\
g_{\tilde f_L \tilde f_R}^\phi & = & -\frac{m_f}{2} (\mu g_3^\phi
- A_f g_4^\phi) \, ,
\label{eq:hsfcouprl}
\end{eqnarray}
with the couplings $g_i^\phi$ listed in Table \ref{tb:hsfcoup}. $I_{3f}$
denotes the third component of the electroweak isospin, $e_f$ the
electric charge of the fermion $f$ and $\sin\theta_W$ the Weinberg
angle, $m_f$ the fermion mass and $M_Z$ the $Z$-boson mass.
\begin{table}[hbt]
\renewcommand{\arraystretch}{1.5}
\begin{center}
\begin{tabular}{|l|c||c|c|c|c|} \hline
$\tilde f$ & $\phi$ & $g^\phi_1$ & $g^\phi_2$ & $g^\phi_3$ & $g^\phi_4$
\\
\hline \hline
& $h$ & $\cos\alpha/\sin\beta$ & $-\sin(\alpha+\beta)$ &
$-\sin\alpha/\sin\beta$ & $\cos\alpha/\sin\beta$ \\
$\tilde u$ & $H$ & $\sin\alpha/\sin\beta$ & $\cos(\alpha+\beta)$ &
$\cos\alpha/\sin\beta$ & $\sin\alpha/\sin\beta$ \\
& $A$ & 0 & 0 & $-1$ & $1/\tgb$ \\ \hline
& $h$ & $-\sin\alpha/\cos\beta$ & $-\sin(\alpha+\beta)$ &
$\cos\alpha/\cos\beta$ & $-\sin\alpha/\cos\beta$ \\
$\tilde d$ & $H$ & $\cos\alpha/\cos\beta$ & $\cos(\alpha+\beta)$ &
$\sin\alpha/\cos\beta$ & $\cos\alpha/\cos\beta$ \\
& $A$ & 0 & 0 & $-1$ & $\tgb$ \\ \hline
\end{tabular}
\renewcommand{\arraystretch}{1.2}
\caption[]{\label{tb:hsfcoup}
\it Coefficients of the neutral MSSM Higgs couplings to sfermion pairs.}
\end{center}
\end{table}

Finally the sfermion couplings to photons are given by \cite{DSUSY}
\begin{equation}
g_{\tilde f_i \tilde f_j}^\gamma = e e_f \delta_{ij}
\end{equation}
and to $Z$ bosons by
\begin{eqnarray}
g_{\tilde f_L \tilde f_L}^Z & = & \frac{e}{s_Wc_W}(I_{3f}-e_f\sin^2\theta_W)
\nonumber \\
g_{\tilde f_R \tilde f_R}^Z & = & -\frac{e}{s_Wc_W} e_f\sin^2\theta_W
\nonumber \\
g_{\tilde f_L \tilde f_R}^Z & = & 0 \, .
\end{eqnarray}
All these couplings have to be rotated to the mass eigenstates by the
mixing angle $\theta_f$. The coupling $e=\sqrt{4\pi\alpha}$ denotes the
elementary electric charge. The Yukawa couplings of squarks to gluinos
and quarks are uniquely determined by the strong coupling $\alpha_s$.
The corresponding Feynman rules can be found in Ref.~\cite{mssmbase}.

Neutral Higgs radiation off top or bottom quarks [$Q=t,b$] in $e^+e^-$
collisions,
\begin{equation}
e^+e^- \to Q\bar{Q} \phi^0 \qquad [\phi^0=h,H,A],
\label{eq:processes}
\end{equation}
is a suitable process for measuring the Yukawa couplings in
supersymmetric theories \cite{dj92}, particularly for the light Higgs
boson $h$ and for moderately heavy Higgs bosons $H$ and $A$. In the
following we present the cross sections for these processes including
the next-to-leading order (NLO) SUSY-QCD corrections. The pure QCD
corrections have been determined before in the SM \cite{di98} and the
MSSM \cite{di00} (approximate results can be found in
Refs.~\cite{da97,da99}). They significantly modify the cross section at
high collider energies.  For SM Higgs boson radiation off top quarks the
full electroweak corrections have recently become available
\cite{tthelw}.  They turn out to be of similar magnitude as the pure QCD
corrections so that an analogous result may be expected for MSSM Higgs
bosons. First results of the SUSY-QCD corrections to $t\bar th$
production appeared in Ref~\cite{zhu}. We will comment on the
large size of these corrections in this paper.

\section{Calculation of NLO SUSY-QCD Corrections}
%        =======================================
In this section we will set up our notation and summarize the present
status of the processes under investigation, before we will describe the
novel SUSY-QCD corrections in detail. \\

\noindent
{\large \bf 2.1 Leading-Order Cross Sections and QCD Corrections} \\
%               ================================================

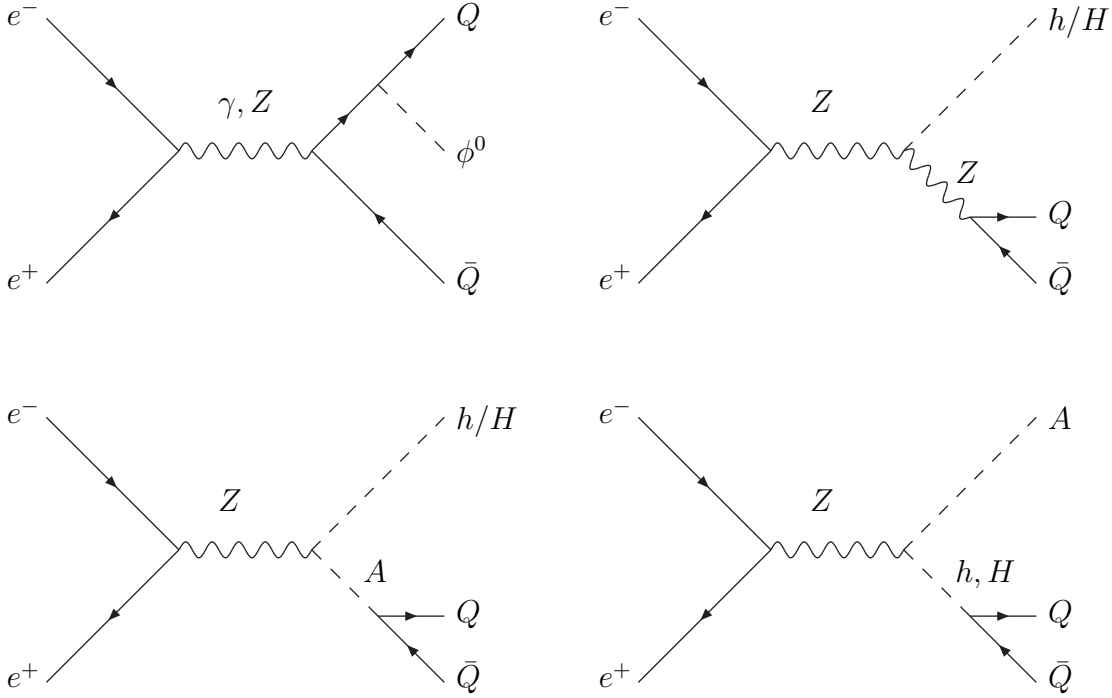
\begin{figure}[hbt]
\begin{picture}(150,110)(-30,0)
\ArrowLine(0,100)(50,50)
\ArrowLine(50,50)(0,0)
\Photon(50,50)(100,50){3}{5}
\ArrowLine(100,50)(125,75)
\ArrowLine(125,75)(150,100)
\ArrowLine(150,0)(100,50)
\DashLine(125,75)(150,50){5}
\put(155,46){$\phi^0$}
\put(-15,98){$e^-$}
\put(-15,-2){$e^+$}
\put(65,65){$\gamma, Z$}
\put(155,98){$Q$}
\put(155,-2){$\bar Q$}
\end{picture}
\begin{picture}(150,110)(-100,0)
\ArrowLine(0,100)(50,50)
\ArrowLine(50,50)(0,0)
\Photon(50,50)(100,50){3}{5}
\Photon(100,50)(125,25){3}{4}
\ArrowLine(125,25)(150,25)
\ArrowLine(150,0)(125,25)
\DashLine(100,50)(150,100){5}
\put(155,96){$h/H$}
\put(-15,98){$e^-$}
\put(-15,-2){$e^+$}
\put(65,65){$Z$}
\put(120,38){$Z$}
\put(155,23){$Q$}
\put(155,-2){$\bar Q$}
\end{picture} \\
\begin{picture}(150,150)(-30,0)
\ArrowLine(0,100)(50,50)
\ArrowLine(50,50)(0,0)
\Photon(50,50)(100,50){3}{5}
\DashLine(100,50)(125,25){5}
\ArrowLine(125,25)(150,25)
\ArrowLine(150,0)(125,25)
\DashLine(100,50)(150,100){5}
\put(155,96){$h/H$}
\put(-15,98){$e^-$}
\put(-15,-2){$e^+$}
\put(65,65){$Z$}
\put(120,38){$A$}
\put(155,23){$Q$}
\put(155,-2){$\bar Q$}
\end{picture}
\begin{picture}(150,150)(-100,0)
\ArrowLine(0,100)(50,50)
\ArrowLine(50,50)(0,0)
\Photon(50,50)(100,50){3}{5}
\DashLine(100,50)(125,25){5}
\ArrowLine(125,25)(150,25)
\ArrowLine(150,0)(125,25)
\DashLine(100,50)(150,100){5}
\put(155,96){$A$}
\put(-15,98){$e^-$}
\put(-15,-2){$e^+$}
\put(65,65){$Z$}
\put(120,38){$h,H$}
\put(155,23){$Q$}
\put(155,-2){$\bar Q$}
\end{picture}
\caption[]{\it \label{fg:lodia} Individual channels of scalar and
pseudoscalar Higgs radiation off heavy quarks $Q=t,b$ in $e^+e^-$
collisions.}
\end{figure}

\noindent
At leading order (LO) Higgs radiation off heavy quarks is described by
the Feynman diagrams of Fig.~\ref{fg:lodia}.
It splits into three different classes of contributions: (i) Higgs
radiation off the heavy (anti)quark, (ii) Higgs radiation off the $Z$
boson (only scalar Higgs radiation) and (iii) $Z$ boson splitting into
scalar-pseudoscalar Higgs pairs with one of them dissociating into a
heavy $Q\bar Q$ pair. Depending on the masses of the corresponding
particles resonant contributions will arise, which require the inclusion
of finite decay widths of the $Z$ and Higgs bosons in the corresponding
propagators. We have used conventional Breit-Wigner propagators for the
resonant $Z\to b\bar b$ and $\phi^0\to t\bar t/b\bar b$ decays as in
previous analyses \cite{di00}. The LO matrix elements can be decomposed
according to the different Higgs couplings,
\bea
{\cal M}_{LO}^{h/H} & = & g_Q^{h/H} {\cal C}^{LO}_1 + g_Z^{h/H} {\cal
C}^{LO}_2 + g_Q^A {\cal C}^{LO}_3 \nonumber \\
{\cal M}_{LO}^A & = & g_Q^A {\cal D}^{LO}_1 + g_Q^h {\cal
D}^{LO}_2 + g_Q^H {\cal D}^{LO}_3 \, .
\eea
The LO cross sections can be cast into the form
\beq
\sigma_{LO} = \int dPS_3 \sum_{spins,colours} \overline{|{\cal
M}_{LO}^{\phi^0}|^2}
\eeq
with $dPS_3$ denoting the three-particle phase space element. The sum
has to be taken over the spins of the initial and final-state fermions
and colours supplemented by an average over the initial $e^+e^-$ spins.

We include the QCD corrections of Ref.~\cite{di00} with the QCD coupling
$\alpha_s$ evaluated at NLO with 5 active flavours at the
renormalization scale $\mu_R=\sqrt{s}$ with $s$ being the $e^+e^-$
c.m.~energy squared. The coupling $\alpha_s$ is normalized to
$\alpha_s(M_Z^2)=0.119$. The bottom Yukawa couplings are computed at the
scale of the corresponding Higgs-momentum flow. This choice absorbs
large logarithmic contributions of the pure QCD corrections \cite{di00}.

The value of the electromagnetic coupling is taken to be $\alpha =
1/128$ and the Weinberg angle to be $s^2_W=0.23$. The mass of the
$Z$~boson is set to $M_Z=91.187$ GeV, and the pole masses of the top and
bottom quarks are set to $m_t=175$ GeV\footnote{The top mass has been
chosen in accordance with the definitions of the Snowmass-benchmark
points of the MSSM \cite{snowmass}.} and $m_b=4.62$ GeV\footnote{This
value for the perturbative pole mass of the bottom quark corresponds in
NLO to an $\overline{\rm MS}$ mass $\overline{m}_b(\overline{m}_b)=4.28$
GeV.}, respectively \cite{di00}. The masses of the MSSM Higgs bosons and
their couplings are related to $\tgb$ and the pseudoscalar Higgs-boson
mass $M_A$. In the relations we use, higher-order corrections up to two
loops in the effective-potential approach are included \cite{ca95}. The
$Z$ boson width has been chosen as $\Gamma_Z=2.49$ GeV, and the Higgs
boson widths have been computed with the program HDECAY \cite{hdecay}. \\

\noindent
{\large \bf 2.2 SUSY-QCD Corrections} \\
%               ====================

\begin{figure}[hbt]
\begin{picture}(150,110)(-30,0)
\ArrowLine(0,100)(50,50)
\ArrowLine(50,50)(0,0)
\Photon(50,50)(100,50){3}{5}
\ArrowLine(150,0)(100,50)
\ArrowLine(100,50)(105,55)
\CArc(115,65)(14,45,225)
\GlueArc(115,65)(14,45,225){3}{4}
\DashCArc(115,65)(14,225,405){3}
\ArrowLine(125,75)(135,85)
\ArrowLine(135,85)(150,100)
\DashLine(135,85)(160,85){3}
\put(165,82){$\phi^0$}
\put(-15,98){$e^-$}
\put(-15,-2){$e^+$}
\put(65,60){$\gamma, Z$}
\put(155,98){$Q$}
\put(155,-2){$\bar Q$}
\put(97,80){$\tilde g$}
\put(130,50){$\tilde Q$}
\end{picture}
\begin{picture}(150,110)(-100,0)
\ArrowLine(0,100)(50,50)
\ArrowLine(50,50)(0,0)
\Photon(50,50)(100,50){3}{5}
\DashLine(100,50)(120,70){3}
\DashLine(100,50)(120,30){3}
\CArc(100,50)(25,-45,45)
\GlueArc(100,50)(25,-45,45){3}{5}
\ArrowLine(120,70)(135,85)
\ArrowLine(135,85)(150,100)
\DashLine(135,85)(160,85){3}
\ArrowLine(150,0)(120,30)
\put(165,82){$\phi^0$}
\put(-15,98){$e^-$}
\put(-15,-2){$e^+$}
\put(65,60){$\gamma, Z$}
\put(155,98){$Q$}
\put(155,-2){$\bar Q$}
\put(100,65){$\tilde Q$}
\put(130,48){$\tilde g$}
\end{picture} \\
\begin{picture}(150,140)(-30,0)
\ArrowLine(0,100)(50,50)
\ArrowLine(50,50)(0,0)
\Photon(50,50)(100,50){3}{5}
\ArrowLine(150,0)(100,50)
\ArrowLine(100,50)(112,62)
\CArc(122,72)(14,45,225)
\GlueArc(122,72)(14,45,225){3}{5}
\ArrowLine(132,82)(150,100)
\DashLine(112,62)(132,62){3}
\DashLine(132,62)(132,82){3}
\DashLine(132,62)(150,44){3}
\put(155,42){$\phi^0$}
\put(-15,98){$e^-$}
\put(-15,-2){$e^+$}
\put(65,60){$\gamma, Z$}
\put(155,98){$Q$}
\put(155,-2){$\bar Q$}
\put(110,95){$\tilde g$}
\put(135,70){$\tilde Q$}
\end{picture}
\begin{picture}(150,140)(-100,0)
\ArrowLine(0,100)(50,50)
\ArrowLine(50,50)(0,0)
\Photon(50,50)(100,50){3}{5}
\DashLine(100,50)(120,70){3}
\DashLine(100,50)(120,30){3}
\DashLine(120,70)(140,50){3}
\Line(120,30)(140,50)
\Gluon(120,30)(140,50){-3}{3}
\DashLine(120,70)(150,100){3}
\ArrowLine(150,0)(120,30)
\ArrowLine(140,50)(160,50)
\put(155,96){$\phi^0$}
\put(-15,98){$e^-$}
\put(-15,-2){$e^+$}
\put(65,60){$\gamma, Z$}
\put(165,48){$Q$}
\put(155,-2){$\bar Q$}
\put(100,65){$\tilde Q$}
\put(135,30){$\tilde g$}
\end{picture} \\
\caption[]{\it \label{fg:sqcddia} Typical diagrams of the NLO SUSY-QCD
corrections to $e^+e^-\to Q\bar Q\phi^0$ [$Q=t,b$] mediated by gluino
$\tilde g$ and squark $\tilde Q=\tilde t, \tilde b$ exchange.}
\end{figure}
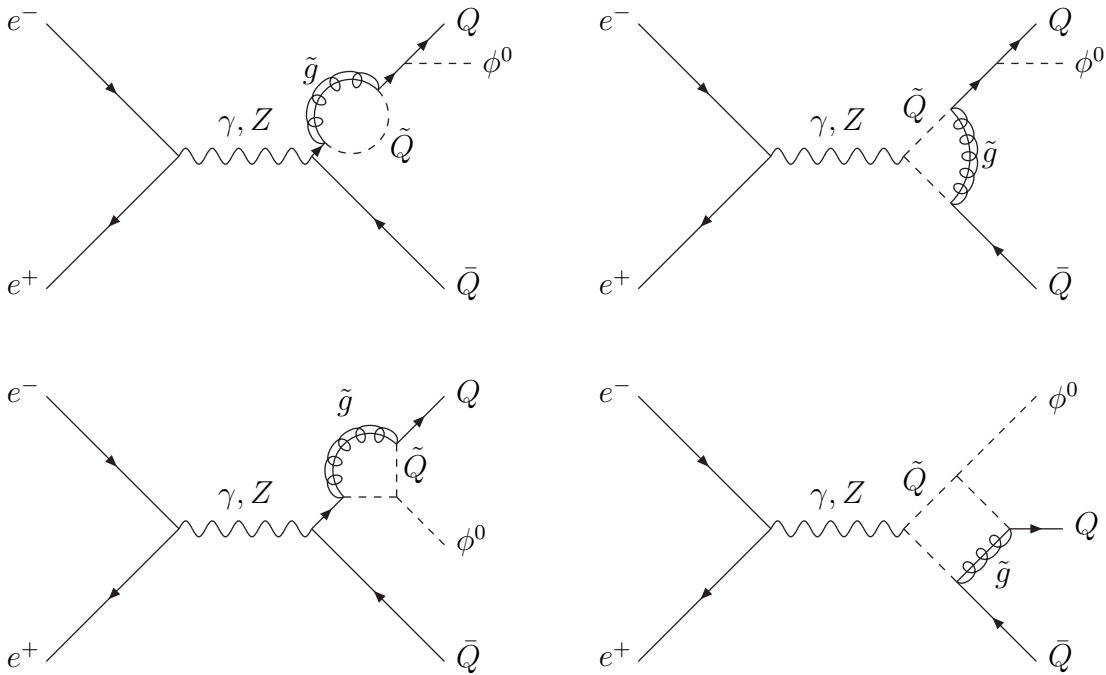

\noindent
The NLO SUSY-QCD corrections arise from virtual gluino and
stop/sbottom exchange contributions as depicted in
Fig.~\ref{fg:sqcddia}.
They consist of self-energy, vertex and box contributions, which are
calculated within dimensional regularization in the standard way.  Since
all virtual particles are massive, no infrared nor collinear
singularities arise. The ultraviolet divergences are removed by the
renormalization of the quark masses and Yukawa couplings. The latter is
connected to the quark mass renormalization. In the case of $t\bar
t$+Higgs production the top mass has been renormalized on-shell in the
propagators as well as in the Yukawa couplings. The same prescription
has also been chosen for the bottom mass, since the virtual gluino and
sbottom masses are too large to develop large logarithmic contributions.
Thus the renormalization of the bottom quark mass is given by
\bea
m_b^0 & = & \overline{m}_b(\mu^2) \left[ 1+\delta_{QCD} + \delta_{SQCD}\right]
\nonumber \\
\delta_{QCD} & = & -\frac{\alpha_s}{\pi} \Gamma(\epsilon) (4\pi)^\epsilon \nonumber \\
\delta_{SQCD} & = & \frac{\tilde \Sigma(m_b)}{\overline{m}_b(\mu^2)} \, ,
\eea
where $m_b^0$ denotes the bare bottom mass, $\overline{m}_b(\mu^2)$ the
$\overline{\rm MS}$ mass at the scale $\mu$ and $\delta_{(S)QCD}$ the
corresponding (SUSY-)QCD counter terms. The SUSY-QCD contribution to the
bottom quark self-energy is displayed by $\tilde \Sigma(m_b)$
with on-shell momentum. This renormalization prescription ensures that
the gluino and sbottom contributions are decoupled from the running of
the bottom Yukawa couplings. Thus we are left with the pure
$\overline{\rm MS}$ Yukawa couplings of QCD.

The final result can be cast into the form
\begin{equation}
\sigma(e^+e^-\to Q\bar Q \phi^0) = \sigma_{LO}(e^+e^-\to Q\bar Q \phi^0)
\left\{ 1+\left[C_{QCD} + C_{SQCD}\right] \frac{\alpha_s(s)}{\pi}
\right\} \, ,
\end{equation}
where $\sigma_{LO}(e^+e^-\to Q\bar Q \phi^0)$ denotes the LO cross
section and $C_{(S)QCD}$ the coefficients of the (SUSY-)QCD corrections.

For large values of $\tgb$ there are significant non-decoupling
corrections to $b\bar b\phi^0$ production, which can be absorbed in the
bottom Yukawa couplings in a universal way \cite{dmb}\footnote{It should
be noted that these contributions vanish for large sbottom and gluino
masses while keeping the $\mu$ parameter fixed \cite{decouple}.
Non-decoupling effects only arise, if the $\mu$ parameter is increased
together with the SUSY particle masses.}. In Refs.~\cite{resum,resum1}
it has been shown that these contributions can be resummed to improve
the reliability of the perturbative result. The resummed bottom Yukawa
couplings are given by\footnote{Analogous effective couplings can be
defined for top quarks, too, but in this case the non-decoupling
contributions are small and thus do not require resummation.}
\bea
\tilde g_b^h & = & \frac{g_b^h}{1+\Delta_b} \left(
1-\frac{\Delta_b}{\tga~\tgb}\right) \nonumber \\
\tilde g_b^H & = & \frac{g_b^H}{1+\Delta_b} \left(
1+\Delta_b \frac{\tga}{\tgb}\right) \nonumber \\
\tilde g_b^A & = & \frac{g_b^A}{1+\Delta_b} \left(
1-\frac{\Delta_b}{\stgb}\right)
\label{eq:dmb}
\eea
with the quantities \cite{resum1}
\bea
\Delta_b & = & \frac{\Delta m_b}{1+\Delta_1} \nonumber \\
\Delta m_b & = & \frac{2}{3}~\frac{\alpha_s}{\pi}~m_{\sgl}~\mu~\tgb~
I(m^2_{\ssb_1},m^2_{\ssb_2},m^2_{\sgl}) \nonumber \\
\Delta_1 & = & -\frac{2}{3}~\frac{\alpha_s}{\pi}~m_{\sgl}~A_b~
I(m^2_{\ssb_1},m^2_{\ssb_2},m^2_{\sgl}) \nonumber \\
I(a,b,c) & = & -\frac{\displaystyle ab\log\frac{a}{b} + bc\log\frac{b}{c}
+ ca\log\frac{c}{a}}{(a-b)(b-c)(c-a)} \; .
\eea
If the LO cross sections are expressed in terms of these resummed
bottom Yukawa couplings, the corresponding NLO pieces have to be
subtracted from the coefficient $C_{SQCD}$ to avoid double counting.
This is equivalent to an additional (finite) renormalization, given
explicitly by
\bea
g_b^\phi & \to & \tilde g_b^\phi \left[ 1 + \Delta_b^\phi \right] +
{\cal O}(\alpha_s^2)
\nonumber \\
\Delta_b^\phi & = & \frac{2}{3}~\frac{\alpha_s}{\pi}~
\kappa_\phi~m_{\sgl}~\mu~\tgb~I(m^2_{\ssb_1},m^2_{\ssb_2},m^2_{\sgl})
\nonumber \\
\kappa_h & = & 1+\frac{1}{\tga~\tgb} \nonumber \\
\kappa_H & = & 1-\frac{\tga}{\tgb} \nonumber \\
\kappa_A & = & 1+\frac{1}{\stgb} \; .
\eea
Thus we have to add additional finite counter terms to the virtual
matrix elements\footnote{Note that in the matrix elements of the
SUSY-QCD corrections we keep the unresummed bottom Yukawa couplings in
order to avoid artificial singularities for vanishing mixing angle
$\alpha$ \cite{resum1}.}
\bea
\Delta {\cal M}_{SQCD}^{h/H} & = & g_b^{h/H} \Delta_b^{h/H} {\cal
C}_1 + g_b^A \Delta_b^A {\cal C}_3
\nonumber \\
\Delta {\cal M}_{SQCD}^A & = & g_b^A \Delta_b^A {\cal D}_1 +
g_b^h \Delta_b^h {\cal D}_2 + g_b^H \Delta_b^H {\cal D}_3 \, .
\label{eq:deltab}
\eea
In the LO matrix elements we use the resummed Yukawa couplings,
\bea
\widetilde{\cal M}_{LO}^{h/H} & = & \tilde g_b^{h/H} {\cal C}^{LO}_1 +
g_Z^{h/H} {\cal C}^{LO}_2 + \tilde g_b^A {\cal C}^{LO}_3 \nonumber \\
\widetilde{\cal M}_{LO}^A & = & \tilde g_b^A {\cal D}^{LO}_1 + \tilde
g_b^h {\cal D}^{LO}_2 + \tilde g_b^H {\cal D}^{LO}_3 \, ,
\eea
so that the SUSY-QCD corrections to the cross sections are given by
\beq
\Delta\sigma_{SQCD} = \sigma_{LO} C_{SQCD} \frac{\alpha_s}{\pi} = \int
dPS_3~2 \Re e \sum_{spins, colours} \overline{\widetilde{\cal
M}_{LO}^{\phi^0\dagger} {\cal M}_{SQCD}^{\phi^0}} \, .
\label{eq:delsig}
\eeq
In the QCD corrections we also insert the resummed bottom Yukawa couplings
everywhere, since the non-decoupling terms $\Delta_b^{\phi^0}$
factorize from the pure QCD corrections initiated by light particle
interactions.

\section{Results}
%        =======
The numerical results will be presented for a linear $e^+e^-$ collider
with c.m.~energy of 1~TeV. We have chosen the Snowmass point SPS5 for
Higgs radiation off top quarks and SPS1b for the bottom quark case
\cite{snowmass}. The MSSM parameters of these two benchmark scenarios
are given by\footnote{We have neglected the corresponding translations
of $\overline{\rm DR}$ masses into $\overline{\rm MS}$ masses, since
they are not relevant for the characterization of the results.} \\

%\noindent
\underline{SPS5:}
\bea
\tgb  & = & 5 \nonumber \\
\mu   & = & 639.8 \gev \nonumber \\
A_t   & = & -1671.4 \gev \nonumber \\
A_b   & = & -905.6 \gev \nonumber \\
m_{\gl} & = & 710.3 \gev \nonumber \\
m_{\sq_L} & = & 535.2 \gev \nonumber \\
m_{\ssb_R} & = & 620.5 \gev \nonumber \\
m_{\st_R} & = & 360.5 \gev
\eea

%\noindent
\underline{SPS1b:}
\bea
\tgb  & = & 30 \nonumber \\
\mu   & = & 495.6 \gev \nonumber \\
A_t   & = & -729.3 \gev \nonumber \\
A_b   & = & -987.4 \gev \nonumber \\
m_{\gl} & = & 916.1 \gev \nonumber \\
m_{\sq_L} & = & 762.5 \gev \nonumber \\
m_{\ssb_R} & = & 780.3 \gev \nonumber \\
m_{\st_R} & = & 670.7 \gev \, .
\eea
The pseudoscalar Higgs mass is left free in both scenarios in order to
scan the corresponding Higgs mass ranges.

\begin{figure}[hbtp]
 \setlength{\unitlength}{1cm}
 \centering
 \begin{picture}(15,18.8)
  \put(2.0, 6.0){\epsfxsize=11cm \epsfbox{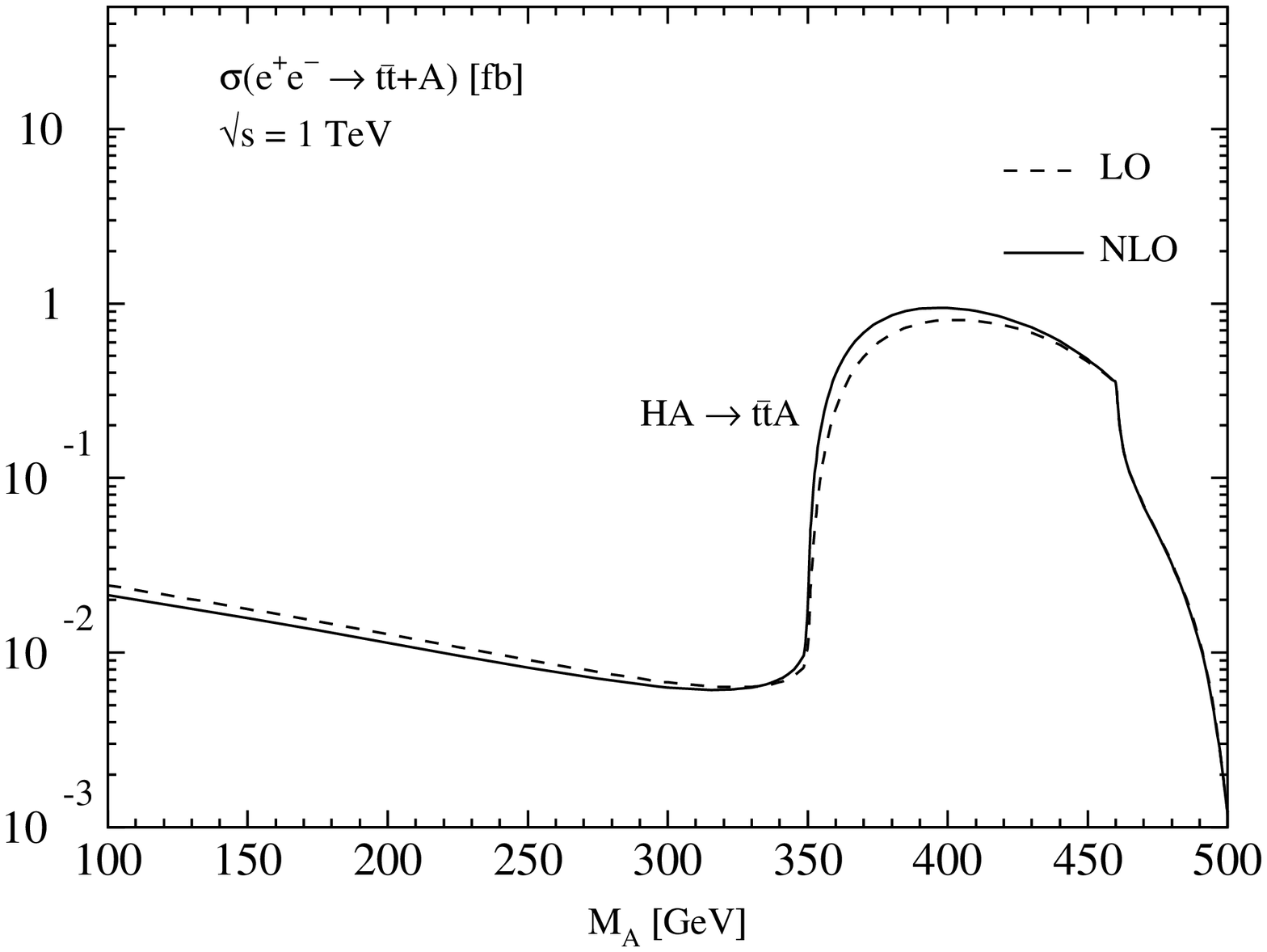}}
  \put(2.0,-3.5){\epsfxsize=11cm \epsfbox{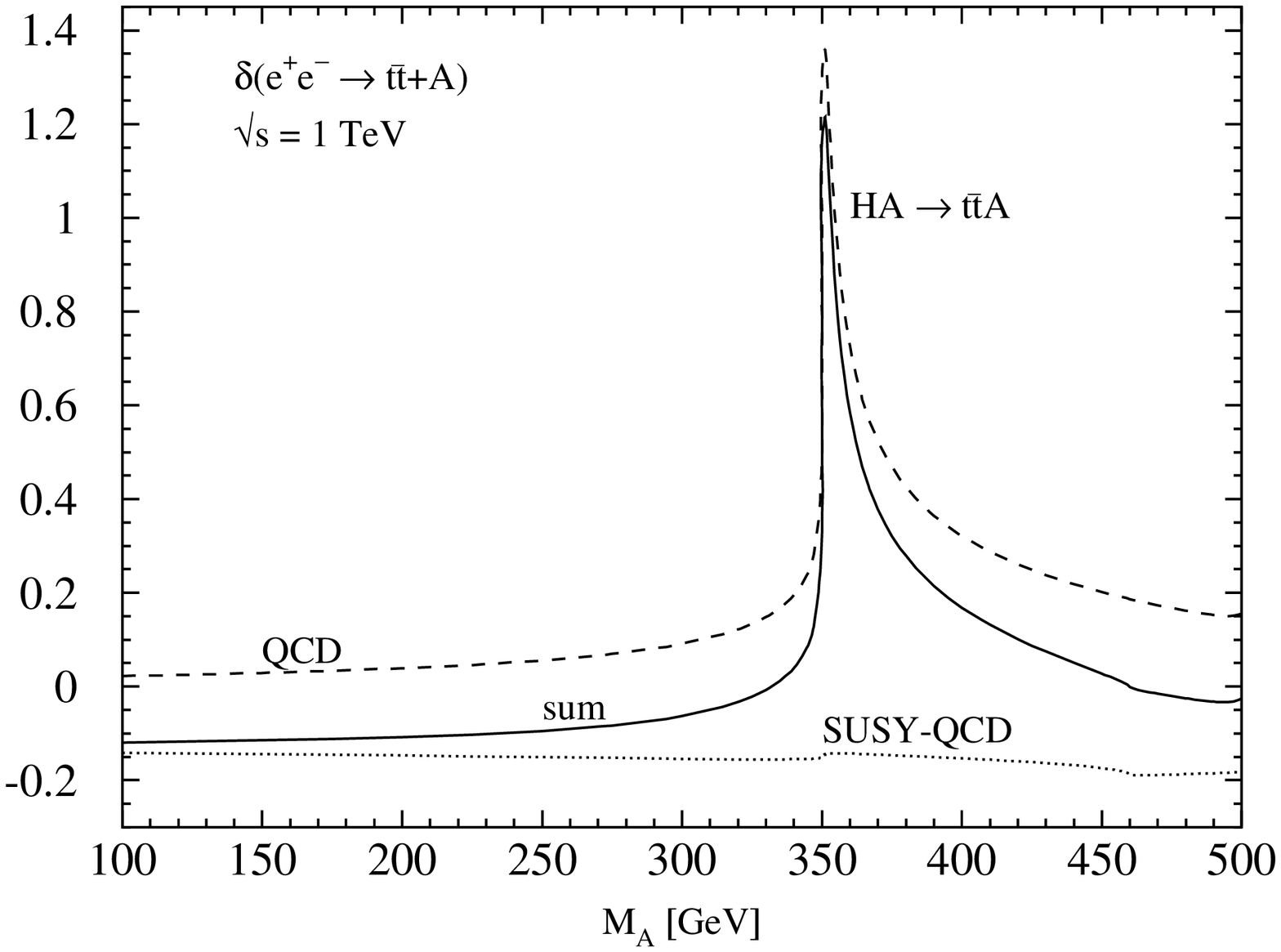}}
  \put(12.3,18.0){\bf (a)}
  \put(12.3, 8.5){\bf (b)}
 \end{picture}
\caption[]{\it \label{fg:tta} (a) Production cross sections of
pseudoscalar Higgs radiation off top quarks in $e^+e^-$ collisions. The
LO cross section is depicted by the dashed line and the full QCD- and
SUSY-QCD corrected cross section by the full line; (b) Relative QCD,
SUSY-QCD and total corrections to pseudoscalar Higgs radiation off top
quarks. The sharp (finite) peak around $M_A=350$ GeV originates from the
Coulomb singularity in the QCD corrections to the resonant $H\to t\bar
t$ decay.}
\end{figure}

\begin{figure}[hbtp]
 \setlength{\unitlength}{1cm}
 \centering
 \begin{picture}(15,18.8)
  \put(2.0, 6.0){\epsfxsize=11cm \epsfbox{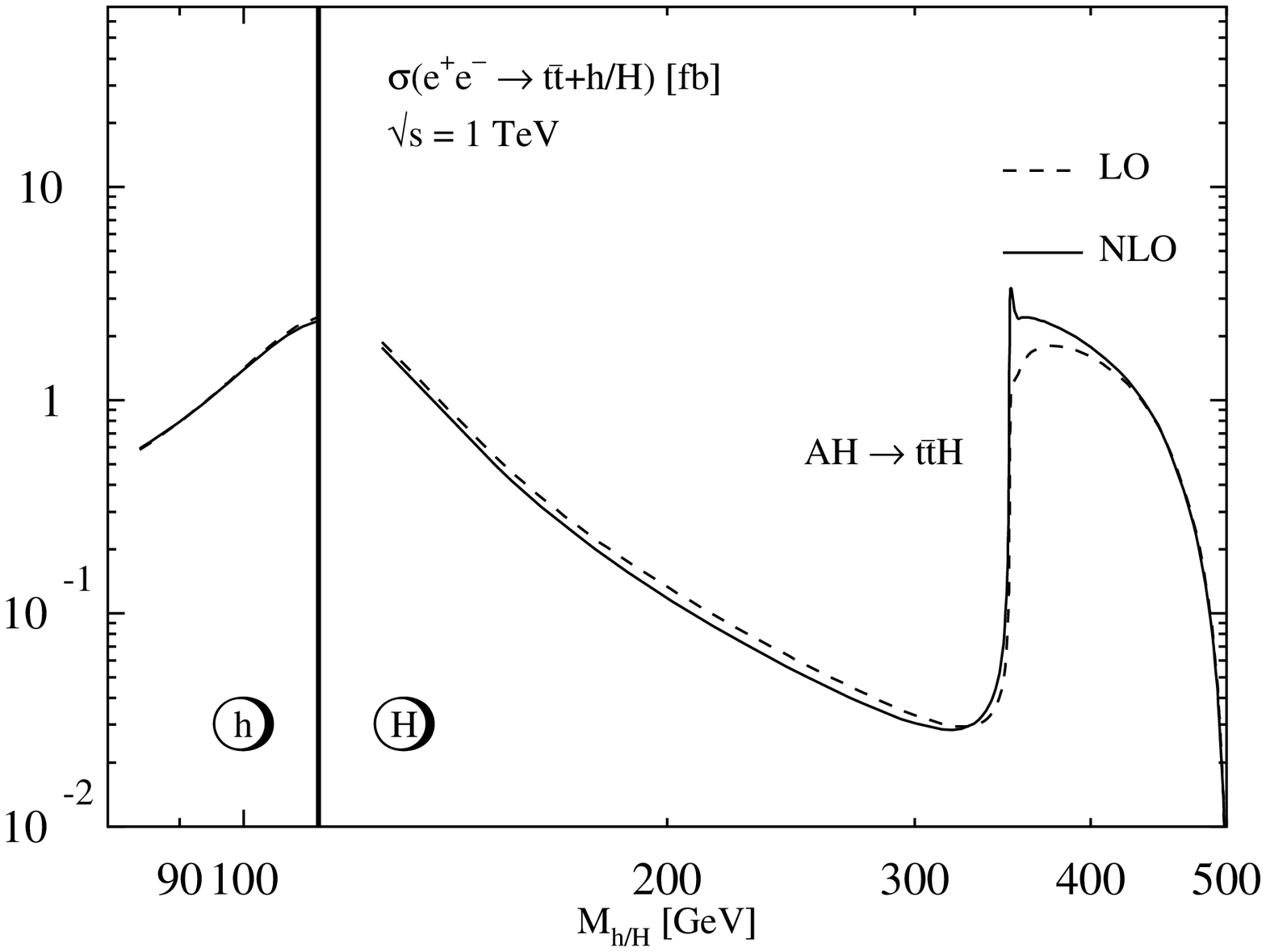}}
  \put(2.0,-3.5){\epsfxsize=11cm \epsfbox{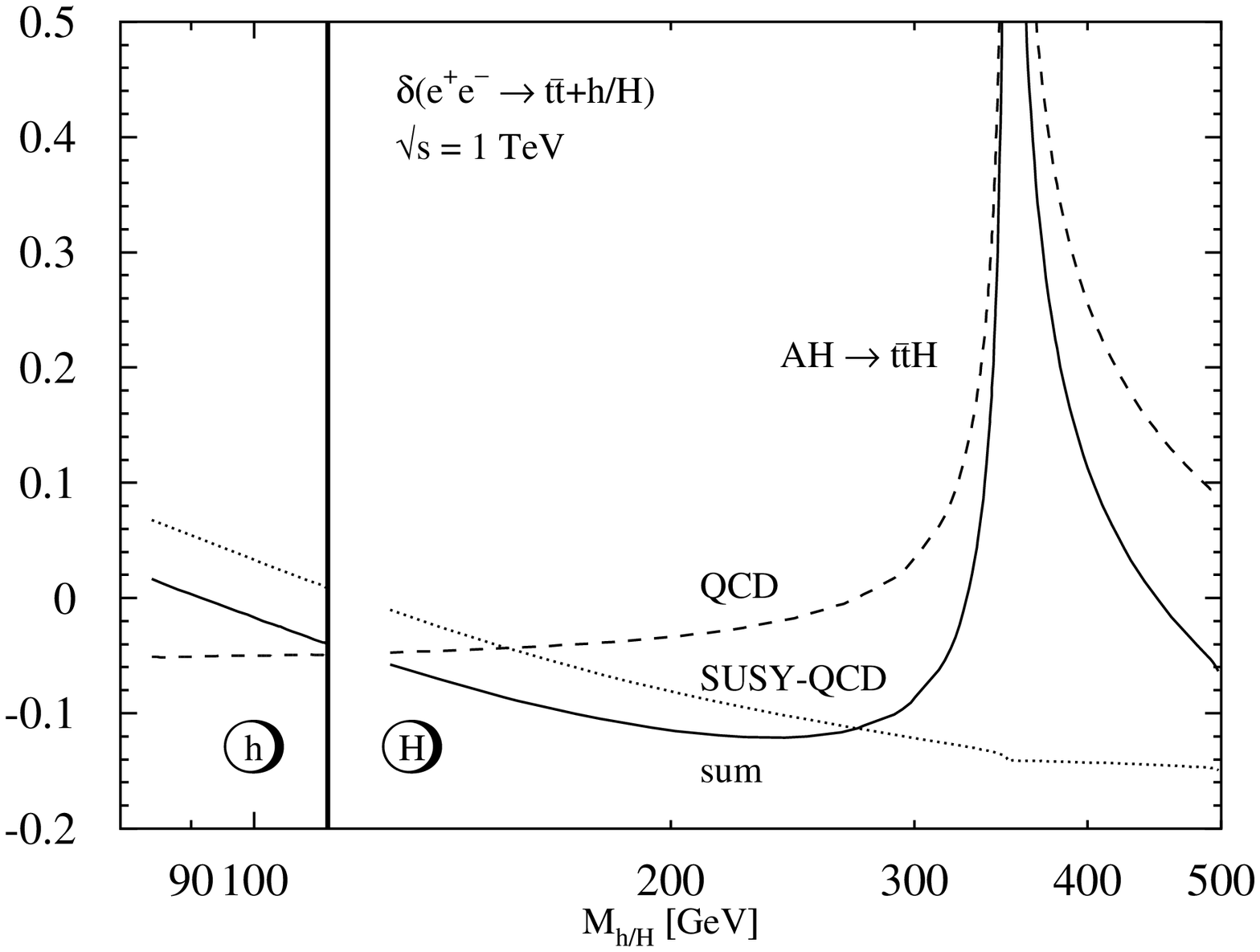}}
  \put(12.3,18.0){\bf (a)}
  \put(12.3, 8.5){\bf (b)}
 \end{picture}
\caption[]{\it \label{fg:tth} (a) Production cross sections of heavy and
light scalar Higgs radiation off top quarks in $e^+e^-$ collisions. The
LO cross section is depicted by the dashed line and the full QCD- and
SUSY-QCD corrected cross section by the full line; (b) Relative QCD,
SUSY-QCD and total corrections to scalar Higgs radiation off top quarks.
The sharp (finite) peak around $M_H=350$ GeV originates from the Coulomb
singularity in the QCD corrections to the resonant $A\to t\bar t$
decay.}
\end{figure}

The total cross section for pseudoscalar Higgs radiation off top quarks
is displayed at LO and NLO in Fig.~\ref{fg:tta}a. The cross section is
small for pseudoscalar Higgs masses below about 350 GeV, while above it
rapidly increases to a level of 1 fb due to the intermediate on-shell
$H\to t\bar t$ decay. The total size of the corrections amounts to about
${\cal O}(10\%)$ apart from the threshold of the resonant contribution,
where the Coulomb singularity raises the QCD corrections to more than
100\% \cite{di98}. The Coulomb singularity is an artefact of the
narrow-width approximation. A proper treatment of the threshold region
requires the inclusion of finite-width effects and QCD-potential
contributions. Thus the result obtained in this work is not valid in a
small margin around the $t\bar t$ threshold of the resonant part.
The individual relative corrections, defined as
$\sigma_{NLO}=\sigma_{LO}(1+\delta_{QCD}+\delta_{SQCD})$, can be
inferred from Fig.~\ref{fg:tta}b. Except for the threshold region of the
resonant part, the QCD corrections are of moderate size \cite{di00}. The
SUSY-QCD corrections are of similar magnitude as the pure QCD
corrections but of opposite sign.  Thus, we observe a large cancellation
of the QCD corrections against the SUSY-QCD part.  This signalizes the
importance of including both types of corrections in future analyses. An
analogous picture emerges for the light and heavy scalar Higgs bosons as
shown in Figs.~\ref{fg:tth}a and b. The cross section for the light
scalar Higgs boson is always of the order of 1 fb with small corrections
due to the partial cancellation of QCD and SUSY-QCD corrections (see
Fig.~\ref{fg:tth}b). The QCD Coulomb singularity for $M_A\sim 350$ GeV
is much more pronounced than in the pseudoscalar case, since for the
heavy scalar Higgs boson the $S$-wave pseudoscalar Higgs decay $A\to
t\bar t$ constitutes the resonant part\footnote{Since the pseudoscalar
Higgs boson $A$ carries the same quantum numbers as the $0^{-+}$ ground
state of the $t\bar t$ pair, the decay $A\to t\bar t$ is dominated by an
$S$-wave contribution at threshold. In contrast the scalar Higgs decay
$H\to t\bar t$ suffers from a $P$-wave suppression at threshold.}. The
relative threshold corrections remain finite in both cases due to the
remaining continuum contributions.

\begin{figure}[hbtp]
 \setlength{\unitlength}{1cm}
 \centering
 \begin{picture}(15,18.8)
  \put(2.0, 6.0){\epsfxsize=11cm \epsfbox{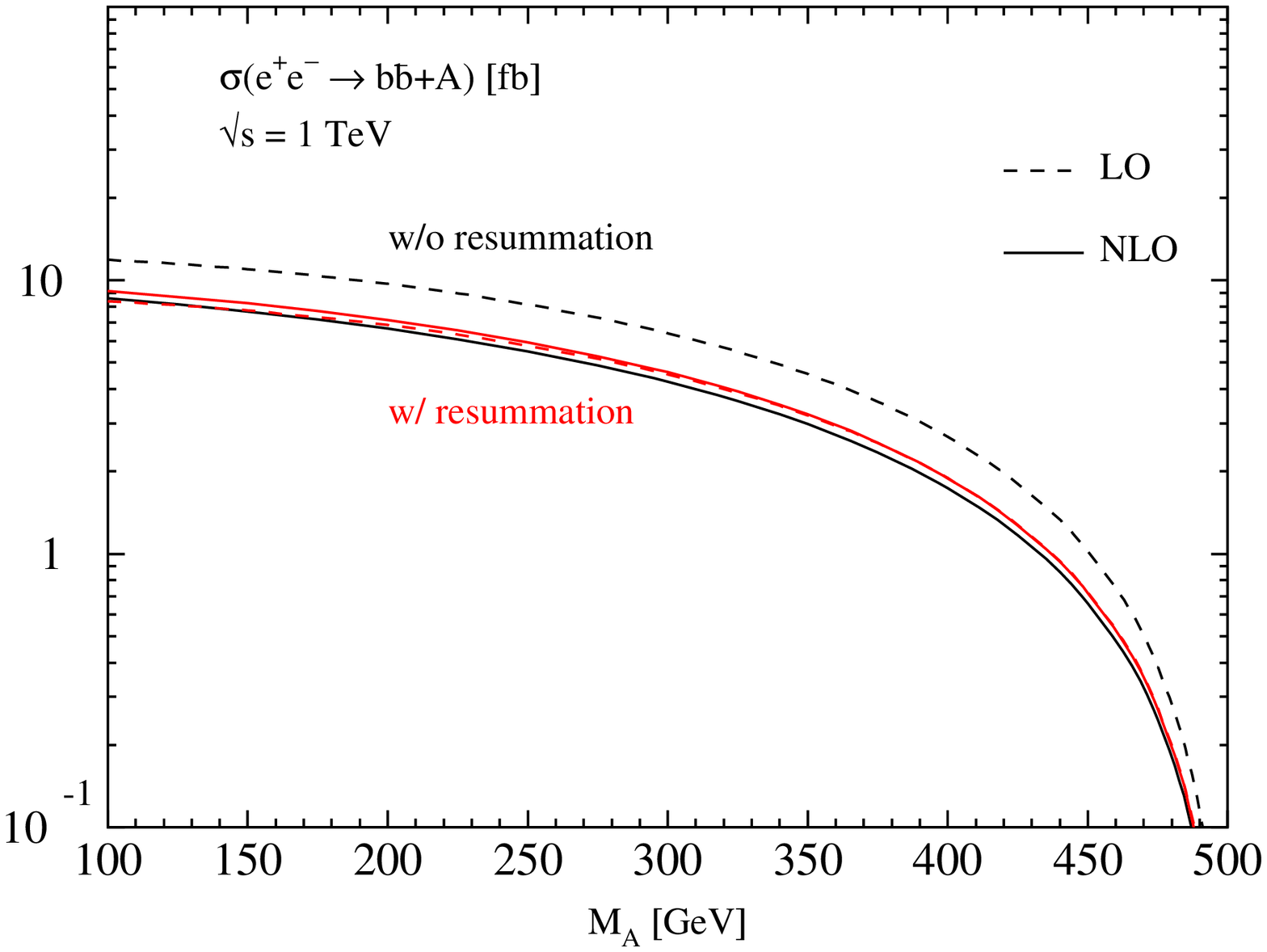}}
  \put(2.0,-3.5){\epsfxsize=11cm \epsfbox{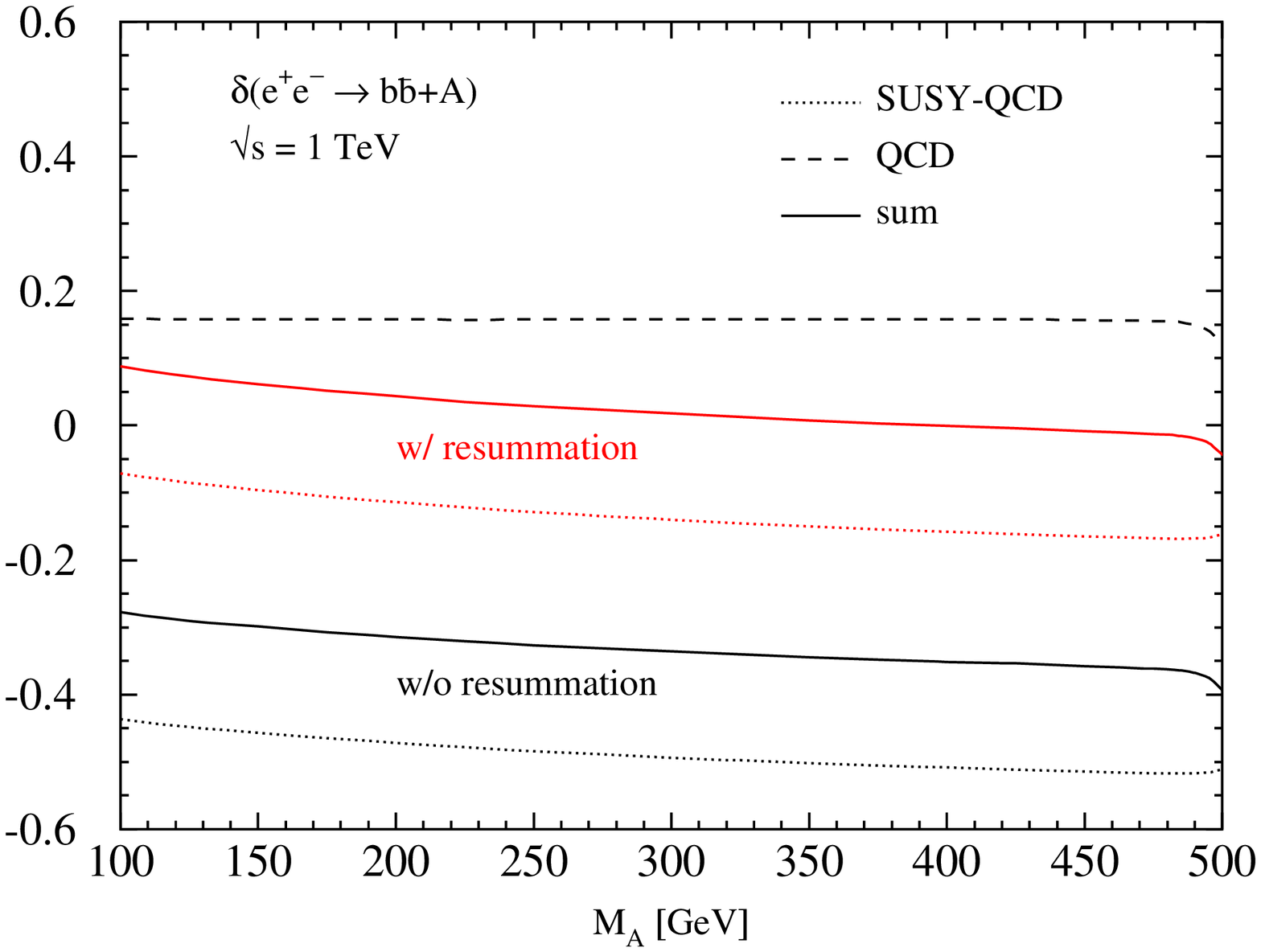}}
  \put(12.3,18.0){\bf (a)}
  \put(12.3, 8.5){\bf (b)}
 \end{picture}
\caption[]{\it \label{fg:bba} (a) Production cross sections of
pseudoscalar Higgs radiation off bottom quarks in $e^+e^-$ collisions with
(red curves) and without (black curves) resummation of the $\Delta_b$
terms. The LO cross section is depicted by the dashed line and the full
QCD- and SUSY-QCD corrected cross section by the full line; (b) Relative
QCD, SUSY-QCD and total corrections to pseudoscalar Higgs radiation off
bottom quarks with (red lines) and without (black lines) resummation. The
pure QCD corrections are indistinguishable in both cases.}
\end{figure}

\begin{figure}[hbtp]
 \setlength{\unitlength}{1cm}
 \centering
 \begin{picture}(15,18.8)
  \put(2.0, 6.0){\epsfxsize=11cm \epsfbox{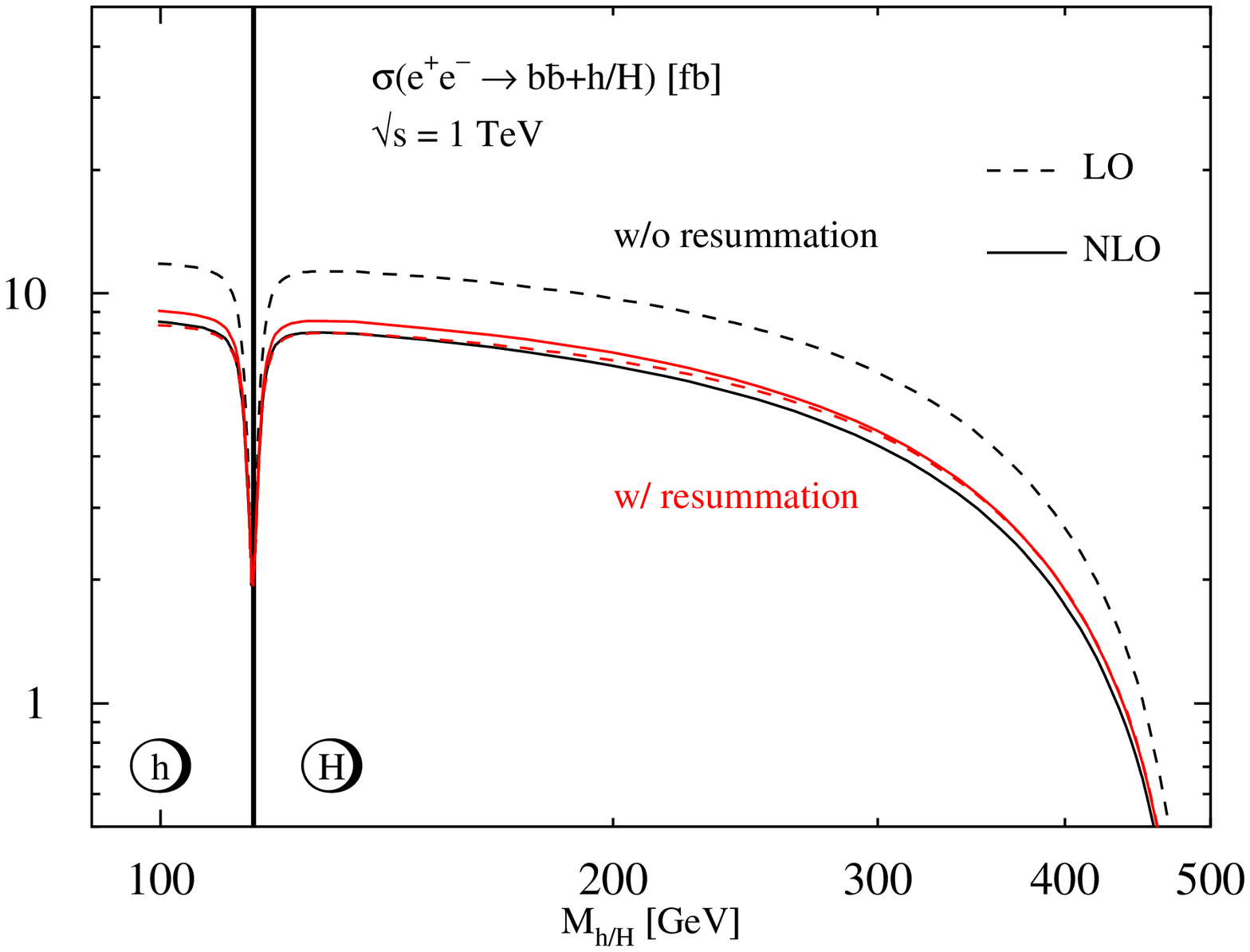}}
  \put(2.0,-3.5){\epsfxsize=11cm \epsfbox{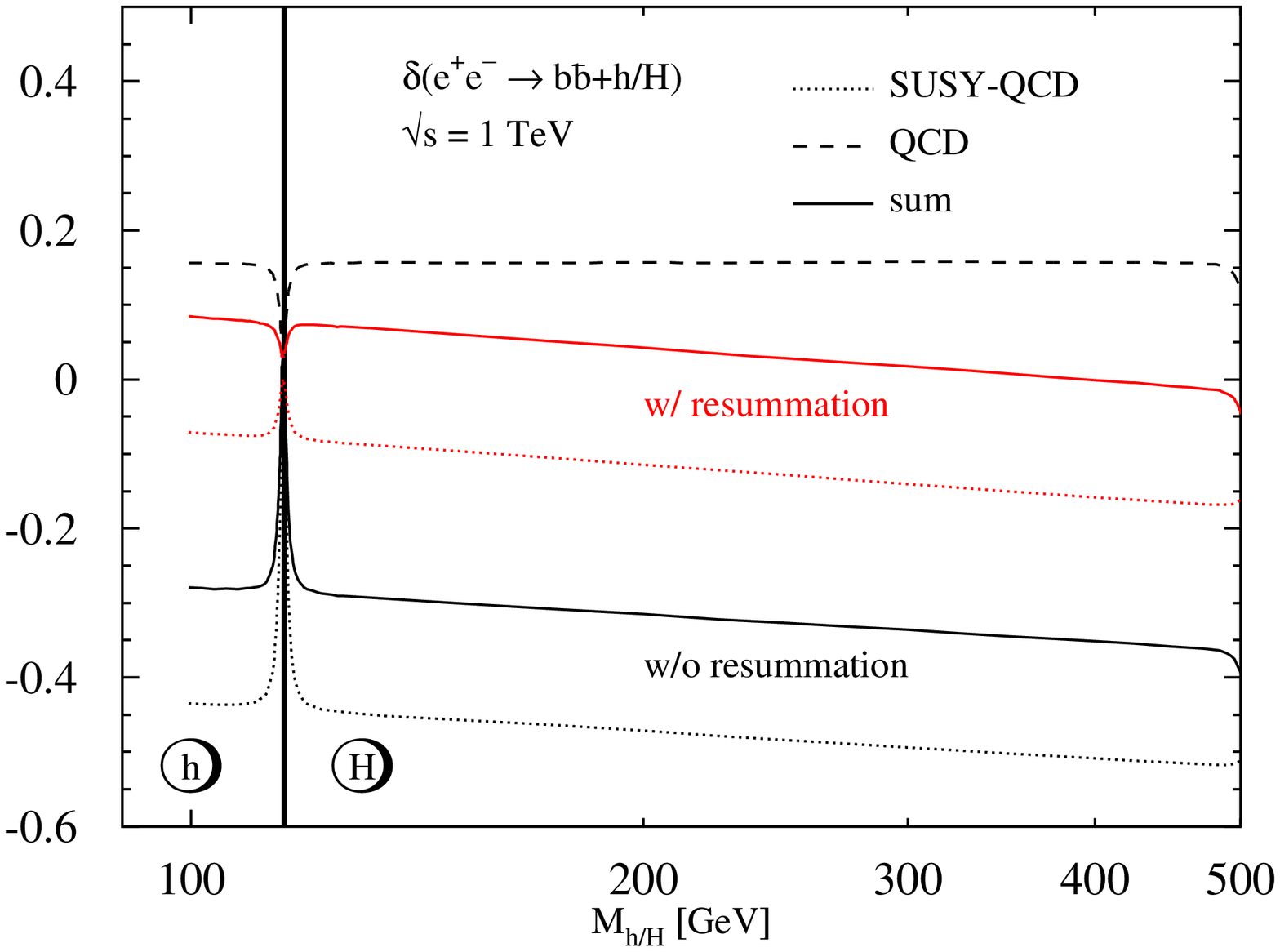}}
  \put(12.3,18.0){\bf (a)}
  \put(12.3, 8.5){\bf (b)}
 \end{picture}
\caption[]{\it \label{fg:bbh} (a) Production cross sections of heavy and
light scalar Higgs radiation off bottom quarks in $e^+e^-$ collisions
with (red curves) and without (black curves) resummation of the
$\Delta_b$ terms. The LO cross section is depicted by the dashed line
and the full QCD- and SUSY-QCD corrected cross section by the full line;
(b) Relative QCD, SUSY-QCD and total corrections to scalar Higgs
radiation off bottom quarks with (red lines) and without (black lines)
resummation. The pure QCD corrections are indistinguishable in both
cases.}
\end{figure}

The results for $b\bar bA$ production are presented in
Fig.~\ref{fg:bba}. The total cross section, shown in Fig.~\ref{fg:bba}a,
reaches a size of ${\cal O}(10~{\rm fb})$ for smaller pseudoscalar
masses. The relative corrections are depicted in Fig.~\ref{fg:bba}b. The
pure SUSY-QCD and total corrections are shown without and with
resummation of the $\Delta_b$ terms according to
Eqs.~(\ref{eq:dmb}--\ref{eq:delsig}). It is clearly visible that the
resummed bottom Yukawa couplings absorb the bulk of the SUSY-QCD
corrections, so that the terms of Eq.~(\ref{eq:deltab}) provide a
reasonable approximation of the final result. After resummation the
SUSY-QCD corrections cancel against the pure QCD corrections to a large
extent. Thus, as in the top quark case the inclusion of both corrections
is of vital importance. A comparison of the total resummed and
unresummed NLO cross sections in Fig.~\ref{fg:bba}a implies good
agreement within 10\% and thus a significant improvement of the
perturbative stability from LO to NLO. An analogous picture emerges for
the light and heavy scalar Higgs bosons as can be inferred from
Fig.~\ref{fg:bbh}. Again the resummed Yukawa couplings absorb the bulk
of the SUSY-QCD corrections. A significant cancellation of the QCD and
SUSY-QCD corrections is observed after resummation in this case, too.
The drops of the relative corrections towards $M_A\sim 500$ GeV in
Figs.~\ref{fg:bba} and \ref{fg:bbh} are caused by the kinematical
closure of the intermediate on-shell $HA$ pair production.

\begin{figure}[hbtp]
 \setlength{\unitlength}{1cm}
 \centering
 \begin{picture}(15,18.5)
  \put(2.0, 5.7){\epsfxsize=11cm \epsfbox{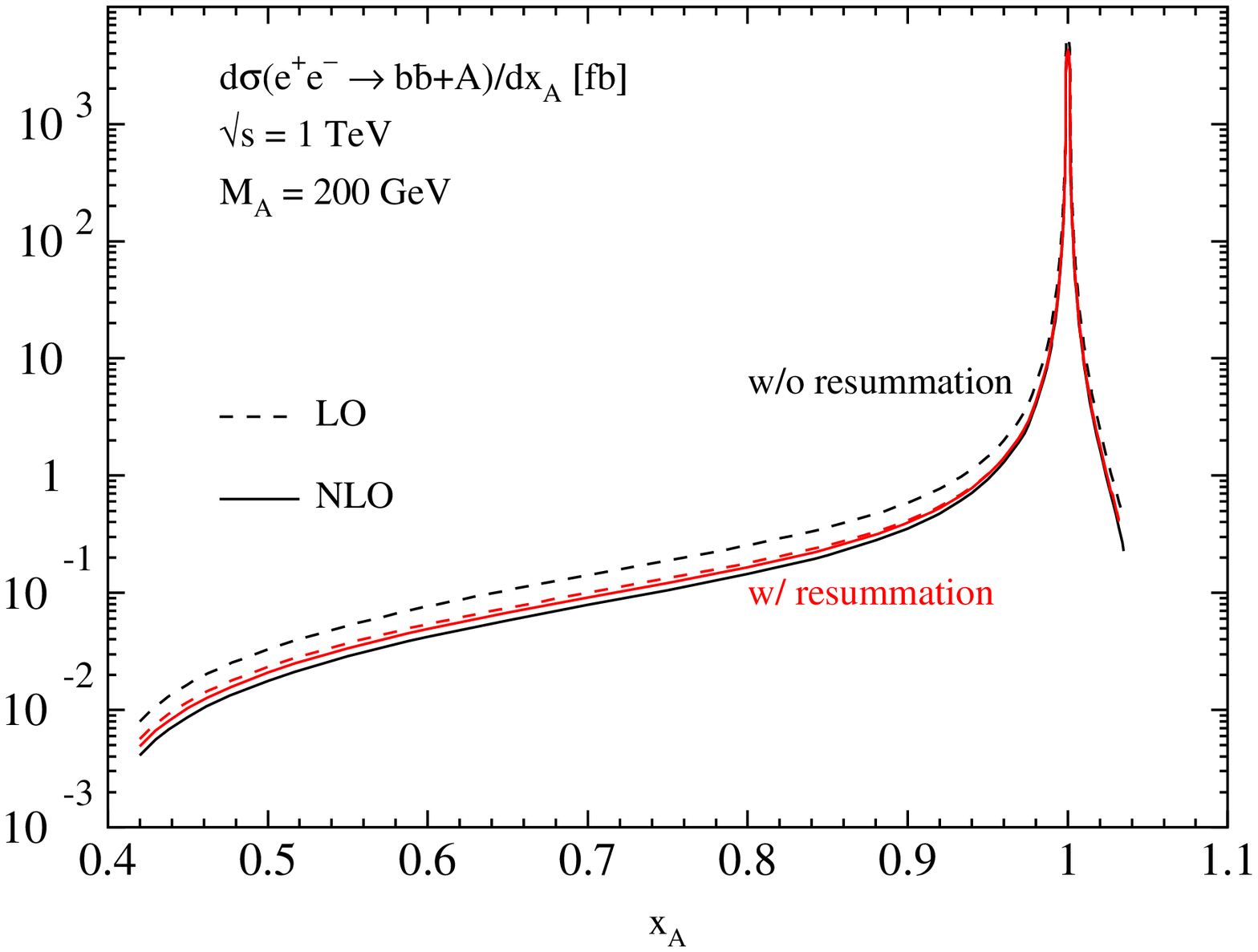}}
  \put(2.0,-3.8){\epsfxsize=11cm \epsfbox{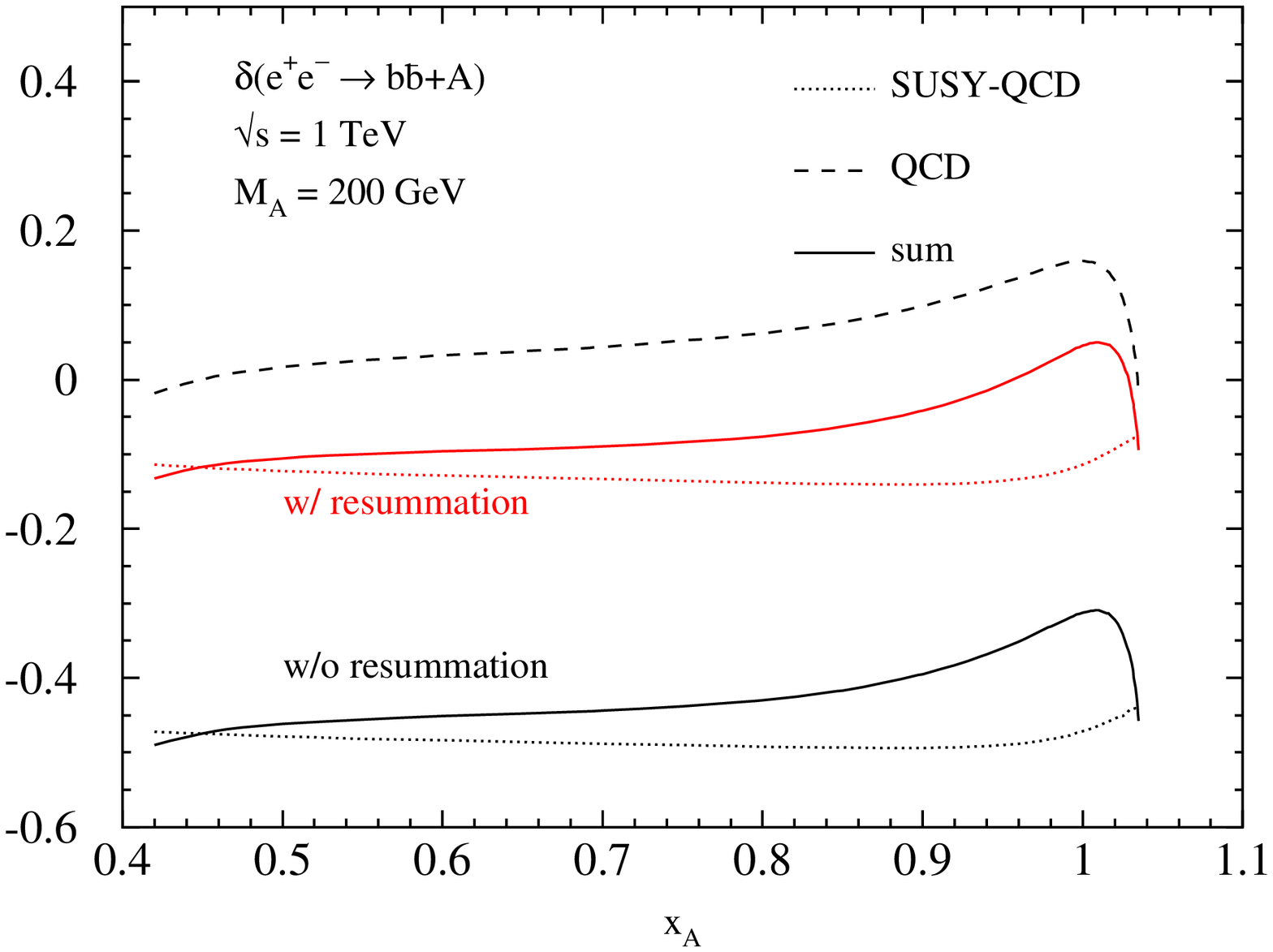}}
  \put(12.3,17.7){\bf (a)}
  \put(12.3, 8.2){\bf (b)}
 \end{picture}
\caption[]{\it \label{fg:xa} (a) Higgs energy distribution of
pseudoscalar Higgs radiation off bottom quarks in $e^+e^-$ collisions
with (red curves) and without (black curves) resummation of the
$\Delta_b$ terms. The LO cross section is depicted by the dashed line
and the full QCD- and SUSY-QCD corrected cross section by the full line.
The peak at $x_A\sim 1$ originates from the resonant $h,H\to b\bar b$
decays; (b) Relative QCD, SUSY-QCD and total corrections to pseudoscalar
Higgs radiation off bottom quarks with (red lines) and without (black
lines) resummation. The pure QCD corrections are indistinguishable in
both cases.}
\end{figure}

Higgs radiation off bottom quarks is, however, dominated by the resonant
$h,H~\to~b\bar b$ decays in the pseudoscalar case and the resonant
$Z,A\to b\bar b$ decays in the scalar case.  Thus, the absorption of the
bulk of the SUSY-QCD part by the resummed Yukawa couplings could be
expected from the analogous findings for the corresponding Higgs decays
\cite{resum1}. In order to investigate, if this also holds for continuum
$b\bar b\phi^0$ production, we analyze the Higgs energy distribution in
Fig.~\ref{fg:xa} for pseudoscalar Higgs radiation off bottom quarks for
a pseudoscalar Higgs mass $M_A=200$ GeV. The dimensionless parameters
$x_{\phi^0}$ are defined as $x_{\phi^0} = 2 E_{\phi^0}/\sqrt{s}$.
Fig.~\ref{fg:xa}a displays the $x_A$ distribution at LO and NLO, while
Fig.~\ref{fg:xa}b exhibits the individual relative corrections.  The
sharp peak at $x_A\sim 1$ originates from the resonant $h,H\to b\bar b$
decays, while the regions apart from the peak represent continuum $b\bar
bA$ production. The resulting picture indeed turns out to be analogous
to the total cross sections.  The bulk of the SUSY-QCD corrections can
be absorbed by the resummed bottom Yukawa couplings leaving moderate
residual corrections. These cancel the pure QCD corrections to a large
extent in the resonant as well as the continuum regions.

\begin{figure}[hbtp]
 \setlength{\unitlength}{1cm}
 \centering
 \begin{picture}(15,18.8)
  \put(2.0, 6.0){\epsfxsize=11cm \epsfbox{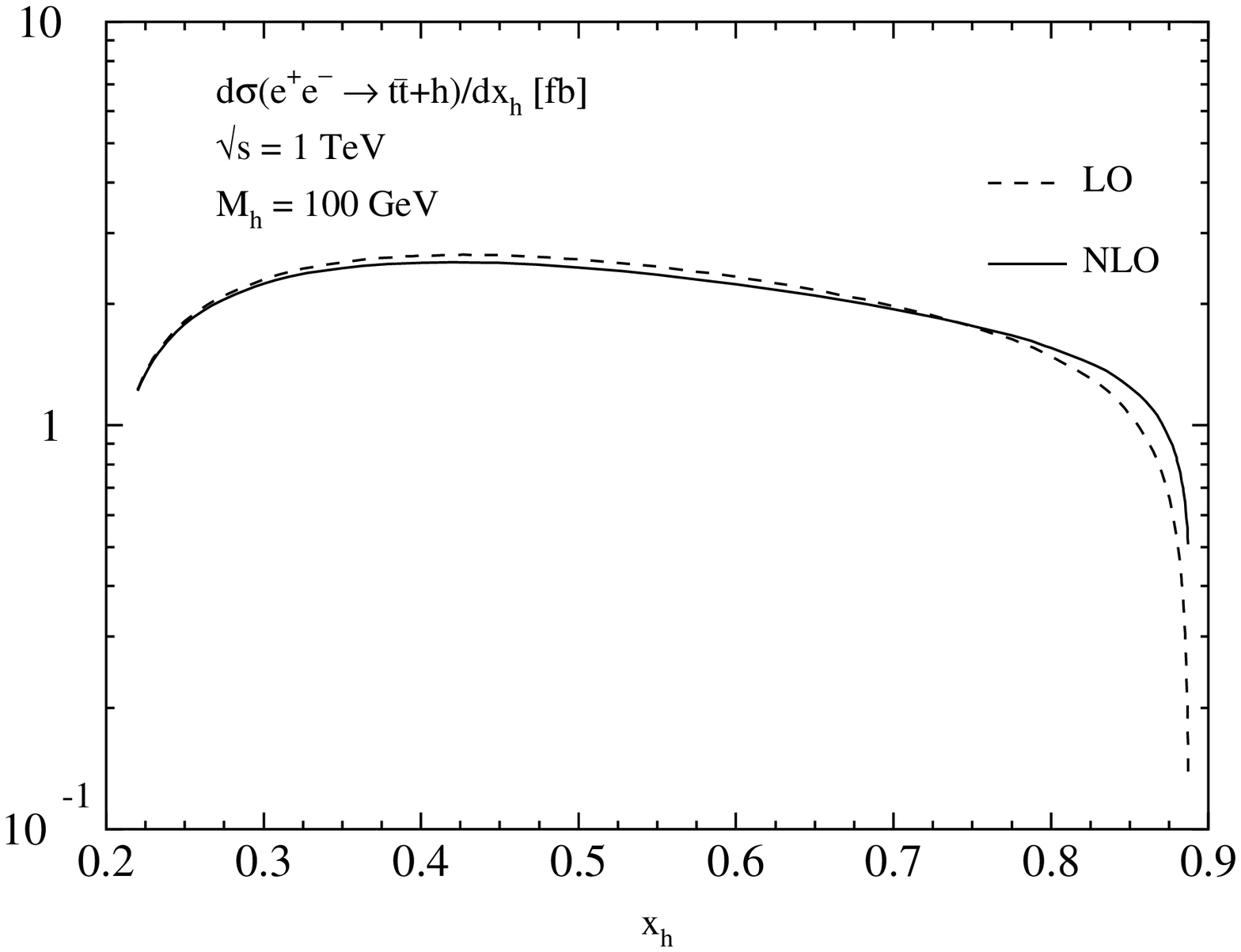}}
  \put(2.0,-3.5){\epsfxsize=11cm \epsfbox{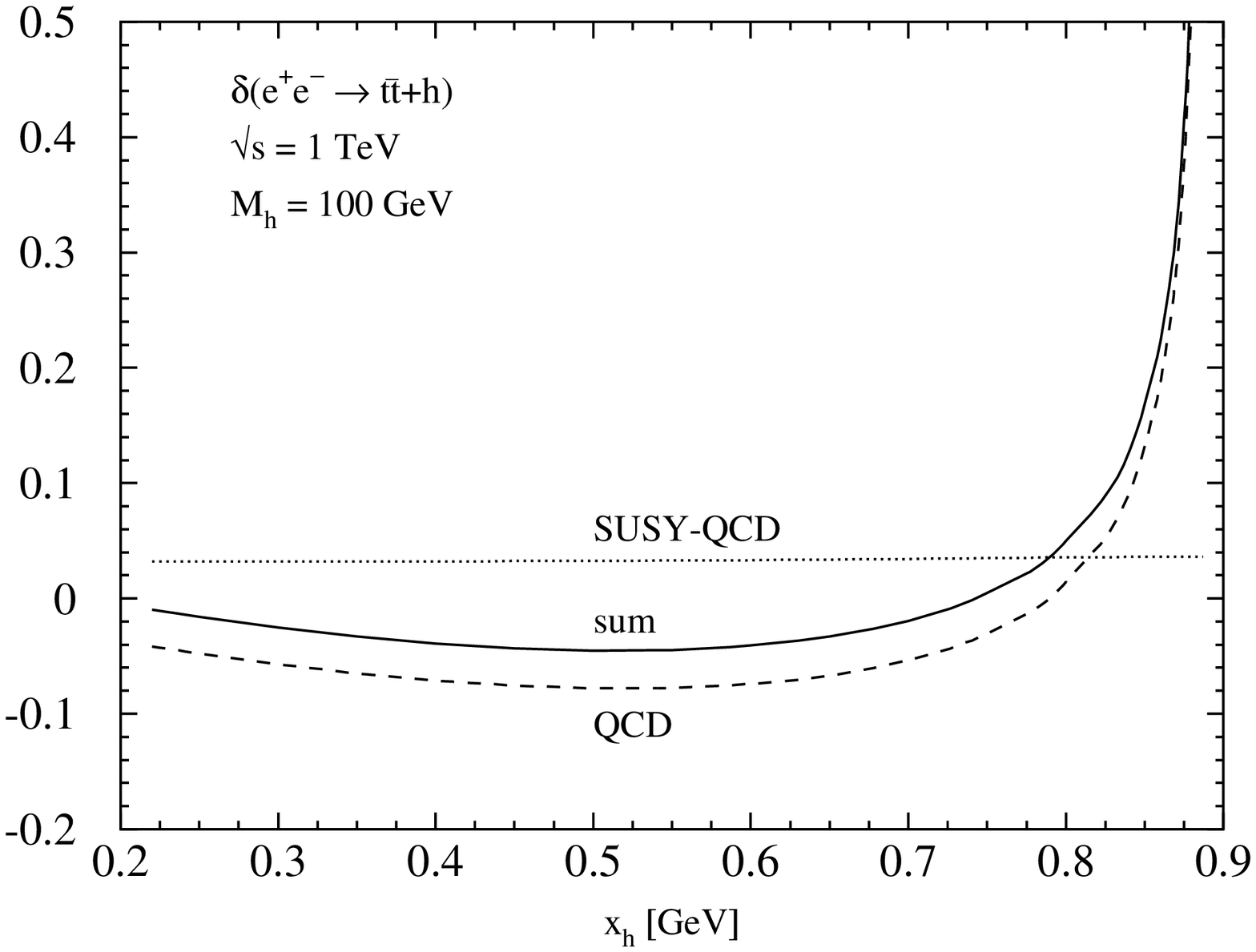}}
  \put(12.2,18.0){\bf (a)}
  \put(12.2, 8.5){\bf (b)}
 \end{picture}
\caption[]{\it \label{fg:xh} (a) Higgs energy distribution of light
scalar Higgs radiation off top quarks in $e^+e^-$ collisions.  The LO
cross section is depicted by the dashed line and the full QCD- and
SUSY-QCD corrected cross section by the full line; (b) Relative QCD,
SUSY-QCD and total corrections to light scalar Higgs radiation off top
quarks. The sharp rise of the QCD corrections towards $x_h\sim 0.9$ is
induced by the Coulomb singularity.}
\end{figure}

The light scalar Higgs energy distribution for Higgs radiation off top
quarks is shown in Fig.~\ref{fg:xh} for a light scalar Higgs mass
$M_h=100$ GeV. For $x_h\lsim 0.8$ both the QCD and SUSY-QCD corrections
are of moderate size and cancel each other partly. The sum of the
corrections amounts to a few per cent. The sharp rise of the QCD
corrections towards $x_h\sim 0.9$ is induced by the Coulomb singularity
at the subthreshold of the $t\bar t$ pair \cite{di00}. It leads to a
finite cross section at the upper bound of the $x_h$ range. Since the
total corrections are not constant, the shape of the Higgs energy
distribution is slightly modified from LO to NLO as can be inferred from
Fig.~\ref{fg:xh}a.

We performed a comparison of our $t\bar th$ results with those of
Ref.~\cite{zhu}. We could not tune our MSSM scenario exactly to the one
of Ref.~\cite{zhu}, since we do not use the same MSSM routines to
calculate the SUSY-particle masses and couplings.  Nevertheless, we have
found reasonable agreement with the results in a scenario very close to
the one in Ref.~\cite{zhu}.  However, this scenario is characterized by
a very extreme choice of parameters. The light stop $\tilde t_1$ is very
light ($m_{\tilde t_1}\sim 100$ GeV), and the gluino mass of 200 GeV is
chosen such that $m_t+M_h\sim m_{\gl}+m_{\st_1}$, so that the thresholds
of the virtual Yukawa vertex corrections in the momentum of the
off-shell (anti)top quarks, which split into $t/\bar t+h$, are very
close to the average momentum flow through the (anti)top propagator (see
the third diagram of Fig.~\ref{fg:sqcddia}). This enhances the size of
the SUSY-QCD corrections considerably. However, the theoretical
uncertainties in the threshold regions are large, and the results cannot
be trusted quantitatively. A better perturbative treatment requires the
inclusion of finite-width effects of the unstable particles as well as
QCD-potential effects. Since the scenario of Ref.~\cite{zhu} belongs to
a very particular region in the MSSM parameter space, this result does
not represent the overall size of the corrections. It can clearly be
inferred from Fig.~6 of Ref.~\cite{zhu} that the corrections are of
moderate size away from this threshold region.

\section{Conclusions}
%        ===========
We have presented the full SUSY-QCD corrections to neutral MSSM Higgs
radiation off top and bottom quarks at linear $e^+e^-$ colliders. The
size of the corrections is of ${\cal O}(10-20\%)$ and of similar
magnitude as the pure QCD corrections obtained in the past, but of
opposite sign, so that significant cancellations occur. This underlines
the relevance of including these corrections in future analyses of these
processes at linear $e^+e^-$ colliders.

At large values of $\tgb$ Higgs radiation off bottom quarks provides a
possibility to measure $\tgb$ \cite{tgb}. In the past it has been
demonstrated that the bulk of the pure QCD corrections can be absorbed
in the running bottom Yukawa couplings, defined at the scale of the
corresponding Higgs momentum flows \cite{di00}. The SUSY-QCD corrections
on the other hand are dominated by the non-decoupling $\Delta m_b$
terms, which can be absorbed and resummed in the corresponding
bottom Yukawa couplings. We have shown that this absorption reduces the
SUSY-QCD corrections to a moderate size, but still cancels the pure QCD
corrections to a large extent in the resonant as well as continuum
regimes.

Based on the size of the electroweak corrections to Standard Model
$t\bar tH$ production the determination of the full SUSY-electroweak
corrections to these processes turns out to be necessary, to reduce
the overall theoretical uncertainties to a level which allows accurate
measurements of the Higgs Yukawa couplings. \\

\noindent
{\bf Acknowledgements.}
We would like to thank S.~Dittmaier, M.~M\"uhlleitner and P.~Zerwas for
valuable comments on the manuscript.

\end{document}